\documentclass[prl,twocolumn,superscriptaddress]{revtex4-2}
\usepackage{amsmath,amssymb,amsfonts,bm,color,graphicx,tabularx,physics}
\usepackage{balance}
\usepackage[unicode=true,colorlinks=true]{hyperref}
\usepackage{bibunits}
 \defaultbibliography{sandwich-junction}
\usepackage[etex=true,export]{adjustbox}

\hypersetup{linkcolor=blue,citecolor=blue,urlcolor=blue}
\usepackage{diagbox}
\usepackage{color}

\newcommand{\beq}{\begin{equation}}
\newcommand{\eeq}{\end{equation}}
\newcommand{\beql}{\begin{equation*}}
\newcommand{\eeql}{\end{equation*}}
\newcommand{\beqn}{\begin{eqnarray}}
\newcommand{\eeqn}{\end{eqnarray}}

\renewcommand{\vec}[1]{\mbox{\boldmath$#1$}}
\begin{document}
\title{Deterministic topological quantum gates for Majorana qubits without ancillary modes}

\author{Su-Qi Zhang}
\affiliation{School of Physics and Institute for Quantum Science and Engineering, Huazhong University of Science and Technology, Wuhan, Hubei 430074, China}
\affiliation{Wuhan National High Magnetic Field Center and Hubei Key Laboratory of Gravitation and Quantum Physics, Wuhan, Hubei 430074, China}

\author{Jian-Song Hong}
\affiliation{International Center for Quantum Materials and School of Physics, Peking University, Beijing 100871, China}

\author{Yuan Xue}
\affiliation{Department of Physics, The University of Texas at Austin, Austin, Texas 78705, USA}

\author{Xun-Jiang Luo}
\affiliation{School of Physics and Institute for Quantum Science and Engineering, Huazhong University of Science and Technology, Wuhan, Hubei 430074, China}
\affiliation{Wuhan National High Magnetic Field Center and Hubei Key Laboratory of Gravitation and Quantum Physics, Wuhan, Hubei 430074, China}

\author{Li-Wei Yu}
\affiliation{Theoretical Physics Division, Chern Institute of Mathematics and LPMC, Nankai University, Tianjin 300071, P. R. China}

\author{Xiong-Jun Liu}
\email{xiongjunliu@pku.edu.cn}
\affiliation{International Center for Quantum Materials and School of Physics, Peking University, Beijing 100871, China}
\affiliation{International Quantum Academy, Shenzhen 518048, China}
\affiliation{CAS Center for Excellence in Topological Quantum Computation, University of Chinese Academy of Sciences, Beijing 100190, China}
\affiliation{Hefei National Laboratory, Hefei 230088, China}

\author{Xin Liu}
\email{phyliuxin@hust.edu.cn}
\affiliation{School of Physics and Institute for Quantum Science and Engineering, Huazhong University of Science and Technology, Wuhan, Hubei 430074, China}
\affiliation{Wuhan National High Magnetic Field Center and Hubei Key Laboratory of Gravitation and Quantum Physics, Wuhan, Hubei 430074, China}
\affiliation{Hefei National Laboratory, Hefei 230088, China}

\date{\today}
	
\begin{abstract}
    The realization of quantum gates in topological quantum computation still confronts significant challenges in both fundamental and practical aspects. Here, we propose a deterministic and fully topologically protected measurement-based scheme to realize the issue of implementing Clifford quantum gates on the Majorana qubits. Our scheme is based on rigorous proof that the single-qubit gate can be performed by leveraging the neighboring Majorana qubit but not disturbing its carried quantum information, eliminating the need for ancillary Majorana zero modes (MZMs) in topological quantum computing. Benefiting from the ancilla-free construction, we show the minimum measurement sequences with four steps to achieve two-qubit Clifford gates by constructing their geometric visualization. To avoid the uncertainty of the measurement-only strategy, we propose manipulating the MZMs in their parameter space to correct the undesired measurement outcomes while maintaining complete topological protection, as demonstrated in a concrete Majorana platform. Our scheme identifies the minimal operations of measurement-based topological and deterministic
    Clifford gates and offers an ancilla-free design of topological quantum computation. 
	\end{abstract}
	\maketitle

Majorana zero modes (MZMs) obey exotic non-Abelian braiding statistics, making them of great interest in fundamental physics and the potential application to topological quantum computation (TQC) \cite{Kitaev2001, Ivanov2001, Nayak2008, Read2000, Motrunich2001}. Braiding two anyons physically by moving one around the other in real space is theoretically straightforward \cite{Alicea2011, Liu2014, Sau2011, Heck2012, Hyart2013, Wu2014, Heck2015, Liu2016} but challenging to implement experimentally \cite{Mourik2012,Deng2012,Rokhinson2012,Das2012,Wang2012,Churchill2013,Xu2014,NadjPerge2014,Chang2015,Sun2016,Albrecht2016,Wiedenmann2016, Bocquillon2016,Jeon2017, Zhang2017, Lutchyn2018,DHong-2018-Sci,FDongLai-2019-CPL,WHaiHu-2018-NatC,Hanaguri-2019-NatM,DHong-2020-Sci,FDongLai-2018-PRX,KLingYuan-2019-NatPhys,DHong-2018-Sci,ZPeng-2019-NatPhys,DHong-2020-NatCom,HJGao2022-nature,DHong2022}. Measurement-based schemes provide an alternative method for generating braiding transformations without physically moving Majorana modes \cite{Bonderson2008, Bonderson2009, Vijay2016, Karzig2017}. However, a comprehensive understanding of all the Clifford gates in these measurement-based schemes is still lacking. In particular, the necessity of ancillary MZMs for implementing quantum gates introduces theoretical, hardware resource, and fabrication complexities. Additionally, the quest for a deterministic measurement-based scheme that combines topological protection and high efficiency remains a significant challenge.

In this letter, we prove the sufficiency and efficiency of using two Majorana qubits, without the need for ancillary MZMs, to implement single- and two-qubit Clifford gates through a measurement-based scheme in a deterministic manner. Firstly, we show a key result that the implementation of a single-qubit gate can be achieved by using a neighboring Majorana qubit but not disturbing its carried quantum information, eliminating the requirement for ancillary MZMs in Majorana-based TQC design. By leveraging the benefits of the ancilla-free construction, we propose the minimal scheme of implementing two-qubit Clifford gates and show rigorously the minimum measurement sequences involve only four steps through a geometric visualization. Further, through a diagrammatic formalism, we demonstrate that Pauli gate \cite{ Bonderson2013, Zheng2015, Zheng2016} can be applied to our scheme to correct undesired outcomes by manipulating the MZMs in their parameter space while maintaining topological protection. Finally, we showcase the experimental accessibility of our proposal by demonstrating its applicability in a concrete Majorana platform.

\textit{Single-qubit gate implementation with two topological qubits-} To perform quantum gate operation with Majorana qubits, the essential task is to exchange two MZMs. The measurement-only method provides a means to braid MZMs without physical movement, but it requires a larger Hilbert space to facilitate the teleportation, rather than the collapse, of quantum information \cite{Bonderson2008, Vijay2016, Karzig2017}. Starting with two Majorana qubits provides an advantage as the eight MZMs offer sufficient redundancy to teleport quantum information initially stored on the computational basis. In contrast, when beginning with only one Majorana qubit, a pair of ancillary MZMs is required to introduce the necessary redundancy. This suggests that we may braid two MZMs in one qubit with the help of the MZMs in the neighbor qubit. However, when attempting to braid two MZMs in one qubit with the help of the neighboring qubit, a crucial question arises: Is it fundamentally allowed without changing the stored quantum information, even if the neighboring qubit can possess arbitrary quantum information and may be entangled with additional qubits? Fortunately, the answer is affirmative, and it can be rigorously proven using the isotopy invariant diagrammatic formalism \cite{Kitaev2006, Bonderson2008, Bonderson2008a, Pachos2012, Simon2021}. 

   \begin{figure} [htbp]
    \centering
    \includegraphics[width= 1\columnwidth]{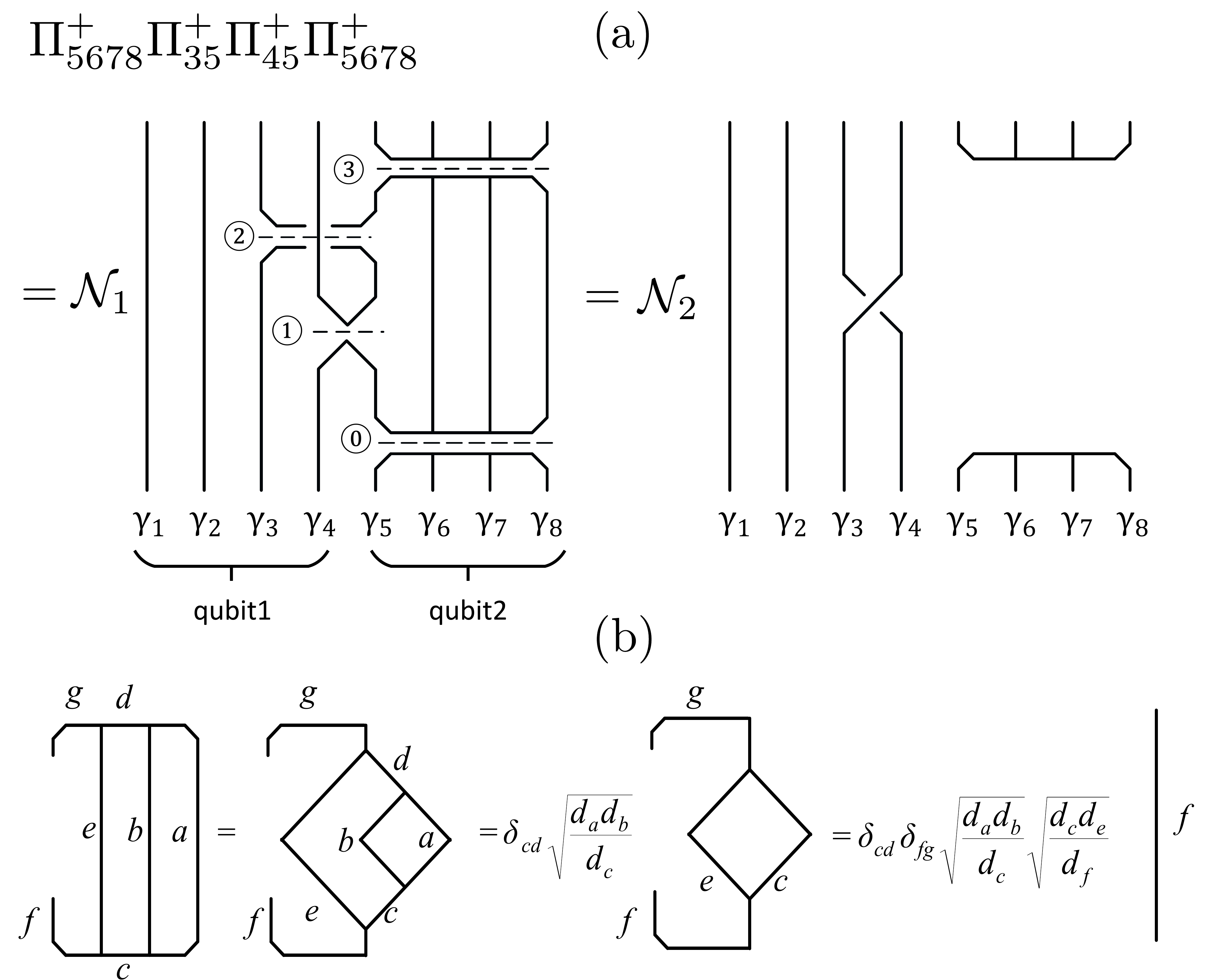}
    \caption{Braiding $\gamma_3$ and $\gamma_4$ within 2 qubits. (a)The time direction is vertical from bottom to top in our representation and $\mathcal{N}_{i}$ is the normalization factor. $\gamma_{i=1-8}$  is the i-th MZM from left to right. The dashed line with the circle j marks the j-th measurement. (b)The locality principle is detailed in (a) for qubit2. The letters $a-g$ are the labels of anyons, $d_a$ is the quantum dimension of $a$, and so forth. }
    \label{Braiding}
    \end{figure}
    
tart from eight Majorana zero modes, the minimal requirement to constructing two Majorana qubits with sparse encoding \cite{Sarma2015}. The first and last four modes form the first and second Majorana qubits, respectively (Fig.~\ref{Braiding}(a)). Throughout our work, the total fermion parity of the eight MZMs remains even. Initially, the quantum information is stored in the computational space where each qubit shares even parity, denoted as $\hat{P}_{1234(5678)} = -\gamma_{1(5)}\gamma_{2(6)}\gamma_{3(7)}\gamma_{4(8)} = +1$. This even parity is guaranteed by the four MZMs measurement, $\Pi^{+}_{1234(5678)}=(1+\hat{P}_{1234(5678)})/2$ (Fig.~\ref{Braiding}(a)). Here and after, we define the fermion parity and measurement operators as $\hat{P}_{i_1...i_n}$ and $\Pi_{i_1...i_n}^{\pm} = (1\pm \hat{P}_{i_1...i_n})/2$ of the $n$ MZMs. Without loss of generality, we first attempt to braid MZMs $\gamma_3$ and $\gamma_4$ in the first qubit with the assistance of the second qubit. The scheme comprises three successive non-destructive topological charge projective measurements ($\Pi^{+}_{45}$, $\Pi^{+}_{35}$ and $\Pi^{+}_{5678}$), as shown in the left-hand side of the diagrammatic representation of Fig.~\ref{Braiding}(a). Note that the measurement operations $\Pi^{+}_{45}$ and $\Pi^{+}_{35}$ do not commute with the fermion parity operators $\hat{P}_{1234}$ and $\hat{P}_{5678}$, respectively, resulting in the teleportation of quantum states out of the computational space. This also reflects in the left-hand side of Fig.~\ref{Braiding}(a) that the strands in both qubit-1 and qubit-2 are involved in the measurement sequences. Remarkably, the strands in qubit-2 can be simplified through the locality principle(Fig.~\ref{Braiding}(b)) into one strand. Meanwhile, the isotopy invariance allows us to freely stretch or slide around a strand so long as its topology remains fixed. After stretching the lines the lines in Fig.~\ref{Braiding}(a), it is clear that the three successive measurements are equivalent to exchange $\gamma_3$ and $\gamma_4$ in qubit-1 and perform identity operation in qubit-2. Thus, we rigorously prove that the exchange of two MZMs in one qubit can be achieved with the assistance of the neighboring qubit but not affecting its quantum information. 
    
    \begin{figure*} [htbp]
    \centering
    \includegraphics[width= 2\columnwidth]{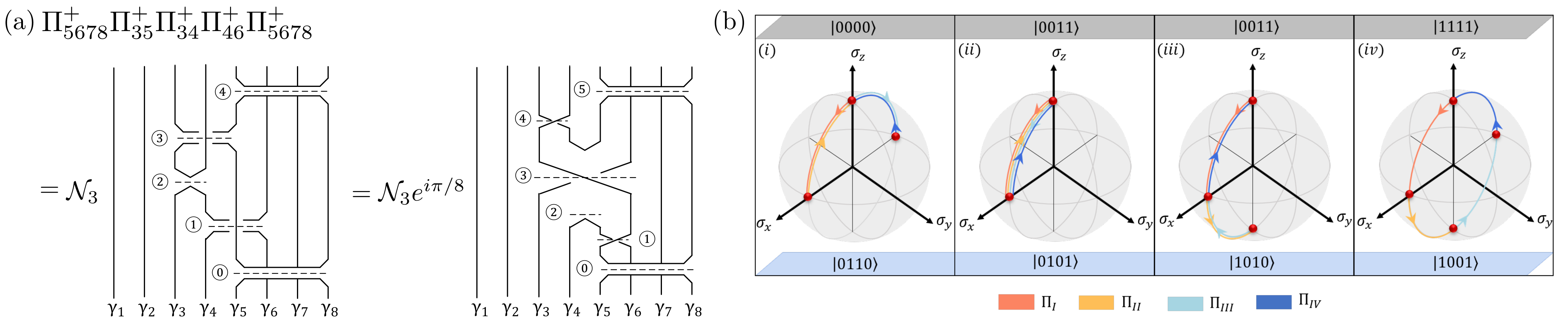}
    \caption{CZ-gate (a) Diagrammatic representation of measurement-based CZ-gate. (b)Geometric visualization of measurement-based CZ-gate in Bloch spheres, with consistent measurement sequences in (a). The upper gray (and lower blue) planes correspond to computational and non-computational spaces, respectively. The colored paths indicate the equivalent measurements. }
    \label{CZ gate}
    \end{figure*}
 
\textit{Minimal measurement-based scheme of Controlled-Z gate -}
Without loss of generality, we first consider the controlled-Z (CZ) gate with the first and second qubits as control and target, respectively. Unlike the $\pi/4$-gate, the CZ gate lacks a standard diagrammatic representation, making it difficult to visualize. For example, we have found that the sequence of four times measurements $\Pi^{+}_{5678}\Pi^{+}_{35}\Pi^{+}_{34}\Pi^{+}_{46}$, can implement the CZ gate in our ancilla-free Majorana qubits through direct calculations. We can make a hindsight demonstration through the continuous deformation of its diagrammatic representation into the celebrated proposal \cite{Sarma2015}, which involves two measurements and three exchange operations in an eight MZMs system (Fig.~\ref{CZ gate}(a)). But this deformation is case by case and lacks a general rule to follow. Recent studies have discovered many measurement sequences to implement the same CZ gate for two Majorana qubits with one or two pairs of ancillary MZMs. It has been shown by brute force that four measurements are the minimum required to realize the CZ gate, but the principles behind these sequences are still unclear \cite{Tran2020}. Furthermore, these measurement sequences involve teleporting quantum states in ten or twelve MZMs Hilbert spaces, which cannot be applied to ancillary-free two Majorana qubits with only eight MZMs. As a result, there is currently no general measurement-based construction method for CZ gates of topological qubits. 
    
To address these issues, we propose a solution based on the ancillary-free two Majorana qubits system. By preserving the even total fermion parity, the measurements teleport the quantum states in eight-dimensional (8D) Hilbert space, which can be decomposed into the direct sum of the four-dimensional computational and non-computational spaces (upper gray plane and lower blue plane in Fig.~\ref{CZ gate}(b)) stabilized by the stabilizers $\{ \hat{P}^{+}_{1-8},\pm \hat{P}^{+}_{5678}\}$ respectively. We define the qubit basis in computational basis as
\begin{align}\label{basis}
       \Psi =\{\psi_{1},\psi_{2},\psi_{3},\psi_{4}\} = \{|0000\rangle , |0011\rangle, |1100\rangle, |1111\rangle \}. \nonumber
\end{align}
To avoid collapsing the quantum information, the first measurement $\Pi^{+}_I$ must teleport the states in a redundant Hilbert space. Therefore, we choose the parity operator $\hat{P}_I$ for the first measurement to anti-commute with $\hat{P}_{5678}$. WLoG, we take $\hat{P}_{I} = \hat{P}_{46}$. The states $\hat{P}_{46} \Psi =\Pi^{+}_{46} \{\psi_{1},\psi_{2},\psi_{3},\psi_{4}\}$ expand the non-computational spaces. To visualize the computational and non-computational Hilbert space, we define the states $\Psi$ and $\hat{P}_{46} \Psi$ as the north and south poles of the four Bloch spheres in which the states $\Pi^{+}_{46} \Psi$ lie along the $+x$ axis in each Bloch sphere (Fig.~\ref{CZ gate}(b)). In this case, the parity operator $\hat{P}_{5678(46)}$ in the four Bloch spheres take the form $\rm{diag}\{\sigma_{z(x)},\sigma_{z(x)},\sigma_{z(x)},\sigma_{z(x)}\}$. Since the CZ gate is diagonal in the computational basis, its measurement sequences should teleport within each Bloch sphere. Consequently, the quantum teleportation through each projective measurement is equivalent to the adiabatic evolution of the states along the $1/4$ great circle connecting the projective points at the Bloch sphere (Fig.~\ref{CZ gate}(b)). Performing the measurement sequences on each qubit basis is equivalent to a unitary evolution in the corresponding Bloch sphere along a closed geodesic, accumulating a geometric phase in each basis. To achieve the CZ-gate, the state $|1111\rangle$ must acquire a $\pi$ phase (Fig.~\ref{CZ gate}(b(iv))), which requires the last Bloch sphere's state teleportation to follow a great circle passing through the north and south poles.
We can choose $\Pi_{II} =(1+\hat{P}_{34})/2$ with $\hat{P}_{34}= \rm{diag}\{\sigma_z,\sigma_z,-\sigma_z,-\sigma_z\}$ and $\Pi_{III} =(1+\hat{P}_{35})/2$ with $\hat{P}_{III}=\rm{diag}\{ -\sigma_x,\sigma_x,\sigma_x,-\sigma_x \}$. Apparently, the projective measurement $\Pi_{\rm II(III)}$ leads the states in different Bloch spheres to undergo different paths, with only the last Bloch sphere's path being a great circle. Therefore, the CZ-gate implemented through the measurements $\Pi_{5678}^{+} \Pi_{35}^{+}  \Pi_{34}^{+} \Pi_{46}^{+}$ without additional ancillary MZMs can be visualized by the paths along the geodesics (Fig.~\ref{CZ gate}(b)). Furthermore, as each teleportation corresponds to the adiabatic evolution of $1/4$ great circle \cite{Karzig2019}, a minimum of four teleportations are needed to implement the CZ gate. We have identified 16 different four-times measurement sequences to implement the CZ gate by varying the choices of $\hat{P}_{\rm I}$. Additionally, we have also determined 8 sequences for the iCZ gate, $\rm{diag}\{1,1,i,-i\}$. These diverse sequences offer flexibility in constructing two-qubit gates for various experimental platforms \cite{supp}.

\begin{figure} [htbp]
\centering
\includegraphics[width= 1\columnwidth]{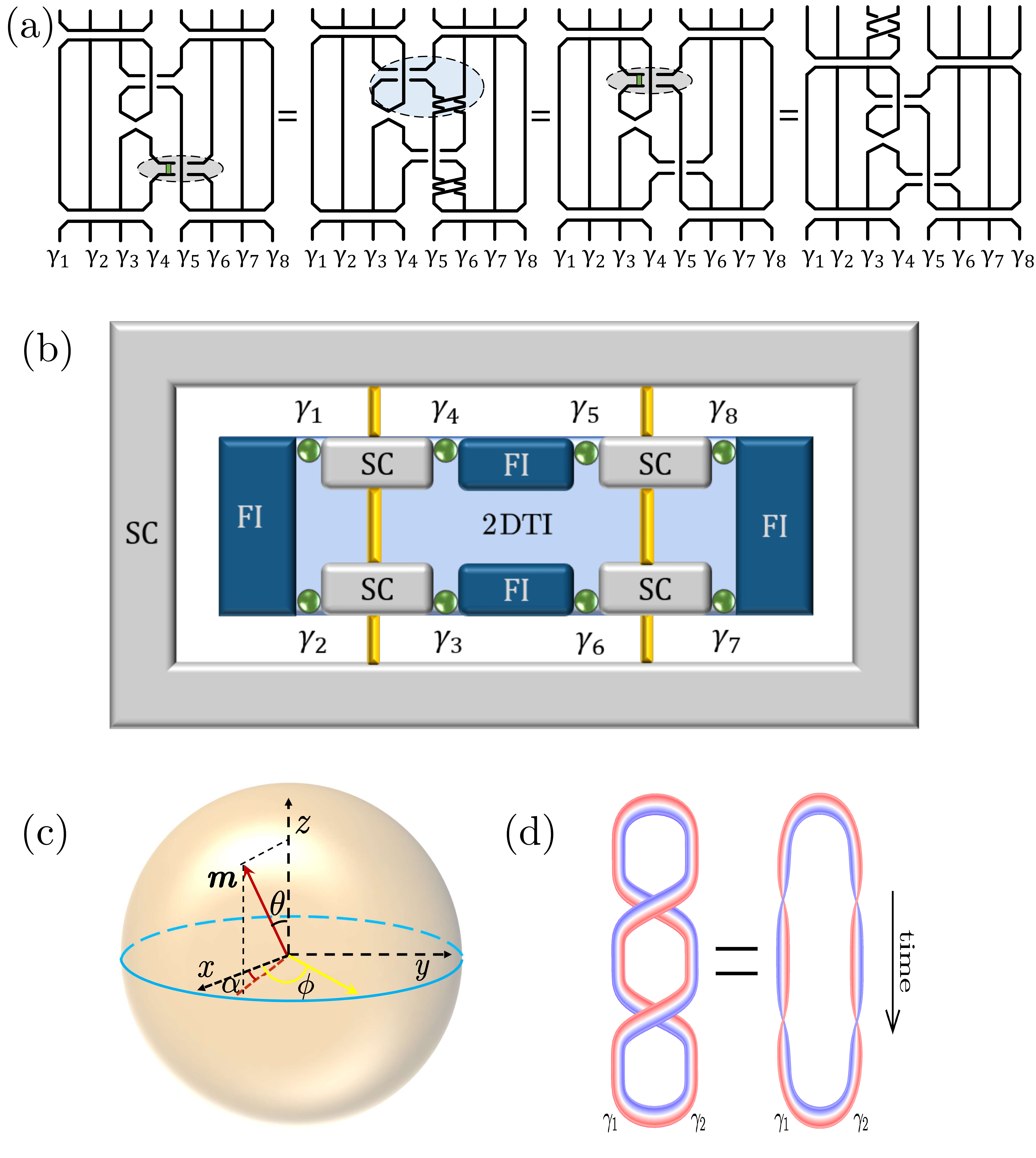}
\caption{(a) A demonstration of propagating undesired outcome in the sequence $\Pi_{5678}^{(+1)} \Pi_{35}^{(+1)} \Pi_{34}^{(+1)} \Pi_{46}^{(-1)} \Pi_{5678}^{(+1)}$ through the full braiding transformation. The small solid green squares mark the measurements as non-vacuum, and dashed gray and blue circles mark the deformations. (b) Two Majorana qubits from the SC/2DTI/FI hybrid system. (c) The relations among the Majorana spin in FI region (yellow arrow), magnetization (red arrow) and SOC field direction ($\bm{e}_z$). The polar and azimuth angles of FI magnetization are $\theta$ and $\alpha$, respectively. The Majorana spin lies in the x-y plane with azimuth angle $\phi$. (d) The equivalence between braiding two MZMs and twisting each Majorana spin by $2\pi$. The arrows indicate the MZM spin.}
\label{Correction}
\end{figure}

\textit{Deterministic topological gates with correction-} It is still a big issue to incorporate the measurement results of non-vacuum into the measurement-based scheme. Although we have the straightforward strategy of forced measurement \cite{Bonderson2008, Vijay2016}, the unknown number and duration of repeated measurements may significantly increase the associated costs. According to the fusion and braiding rules, the different fusion channels can be connected using the following equation
    \beqn\label{eq:fusion_braiding}
    \centering
    \includegraphics[width=0.7\columnwidth]{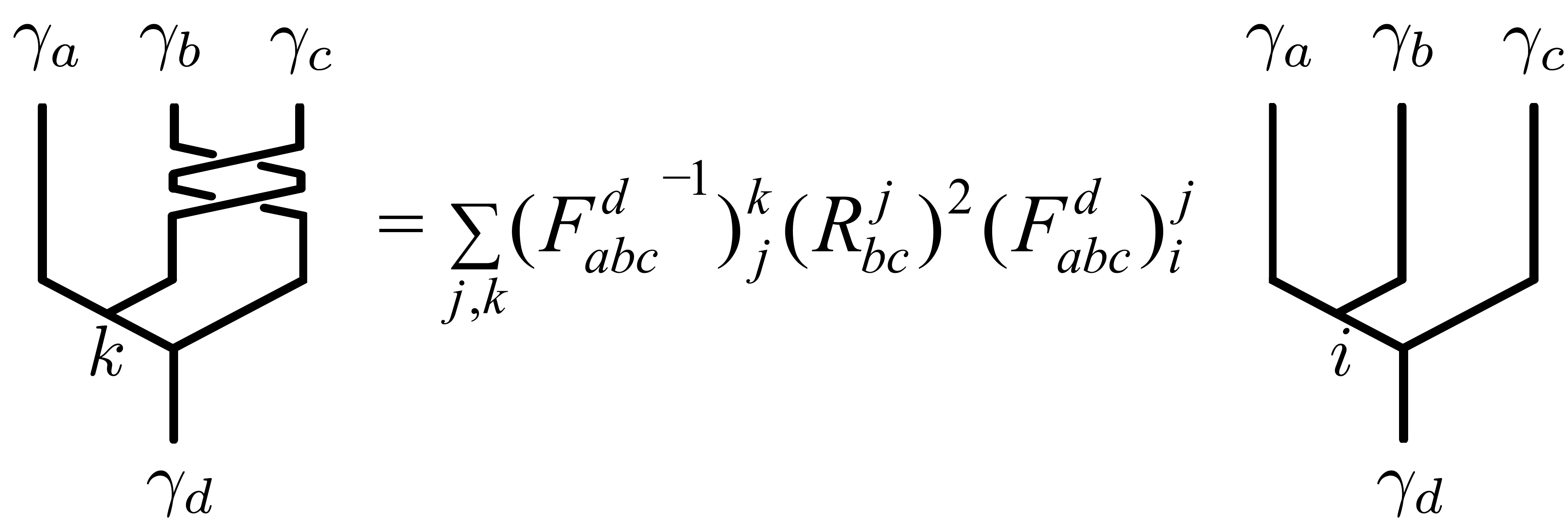}
    \eeqn
where $(F_{abc}^{d})^i_j$ and $R_{ab}^j$ are the fusion and exchange matrices respectively and $a(b,c,d)$ represent the non-Abelian anyons and $i(j,k)$ their fusion channels.
The full braiding of two $Z_{2m}$ parafermions, according to the spin-statistics theorem \cite{Lindner2012, Cheng2012}, follows 
$R^2 \propto e^{i s n^2 \pi/m}$ with $s\in \mathbb{Z}$ , $n$ corresponds to the topological charge and $2m$ the number of fusion channels. For Ising anyons ($s=1$ and $m=1$) with two fusion channels (vacuum and fermion), denoted by $n=0$ and $n=1$ respectively, the pentagon and hexagon identities yield $\sum_j(F_{abc}^{d} {}^{-1})^k_j(R_{ab}^{j})^2 (F_{abc}^{d})^j_i = e^{-i\pi/4}\sigma^x_{ik}$, where $\sigma^x$ acts on the fusion space of $\gamma_a$ and $\gamma_b$. Simplifying Equation \eqref{eq:fusion_braiding}, we have:
 \begin{align}
    \includegraphics[width=0.45\columnwidth]{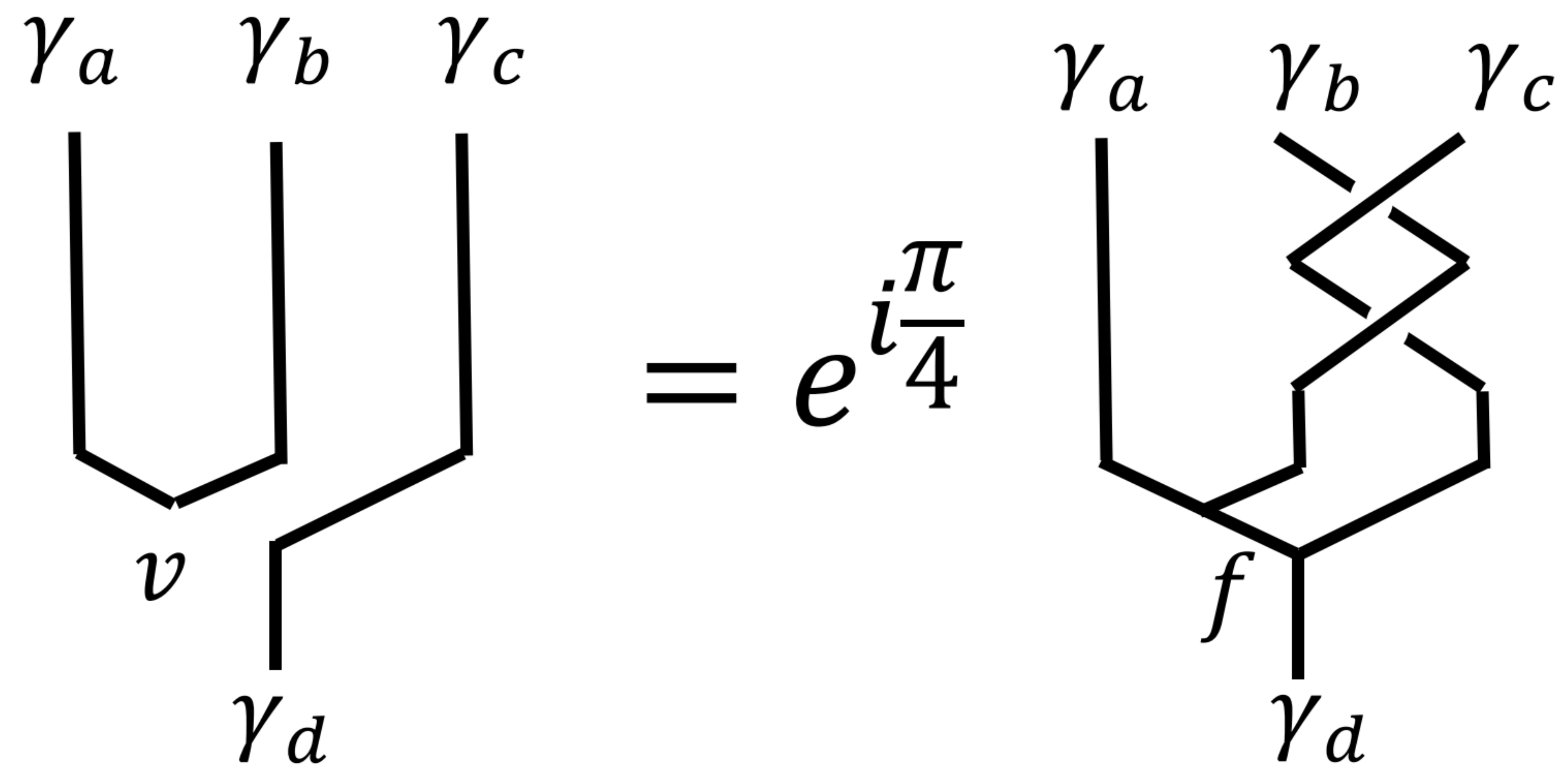}, 
    \includegraphics[width=0.45\columnwidth]{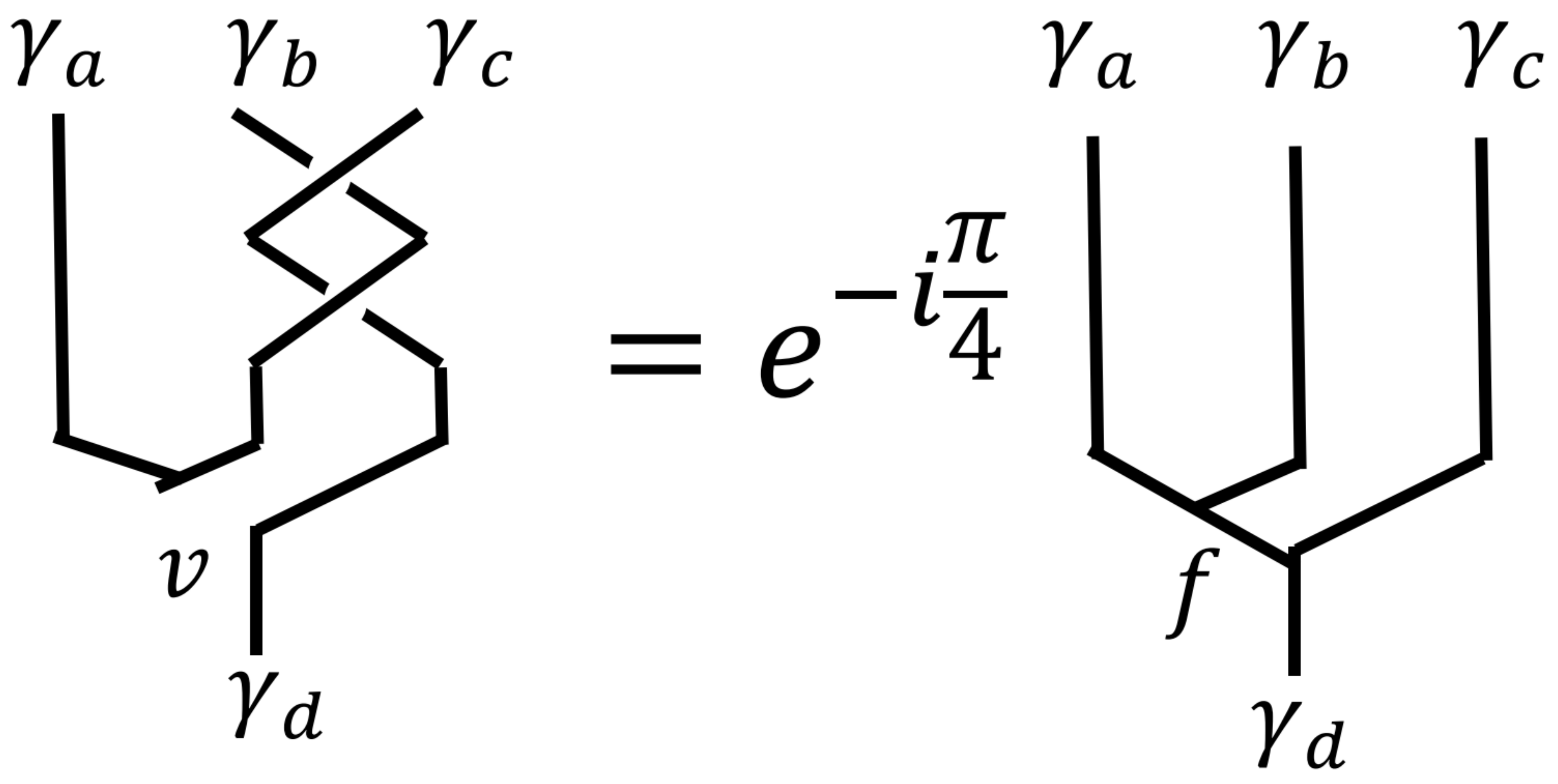},
 \end{align}   
 with $v$ and $f$ the fusion channels of vacuum and fermion. Note that each diagram in the above is a quantum state. Thus the corresponding projector operator satisfies 
 \begin{subequations}
    \begin{tabularx}{\hsize}{@{}XXX@{}}
        \begin{equation}
            \label{Tran-3}
            \includegraphics[width=0.35\columnwidth]{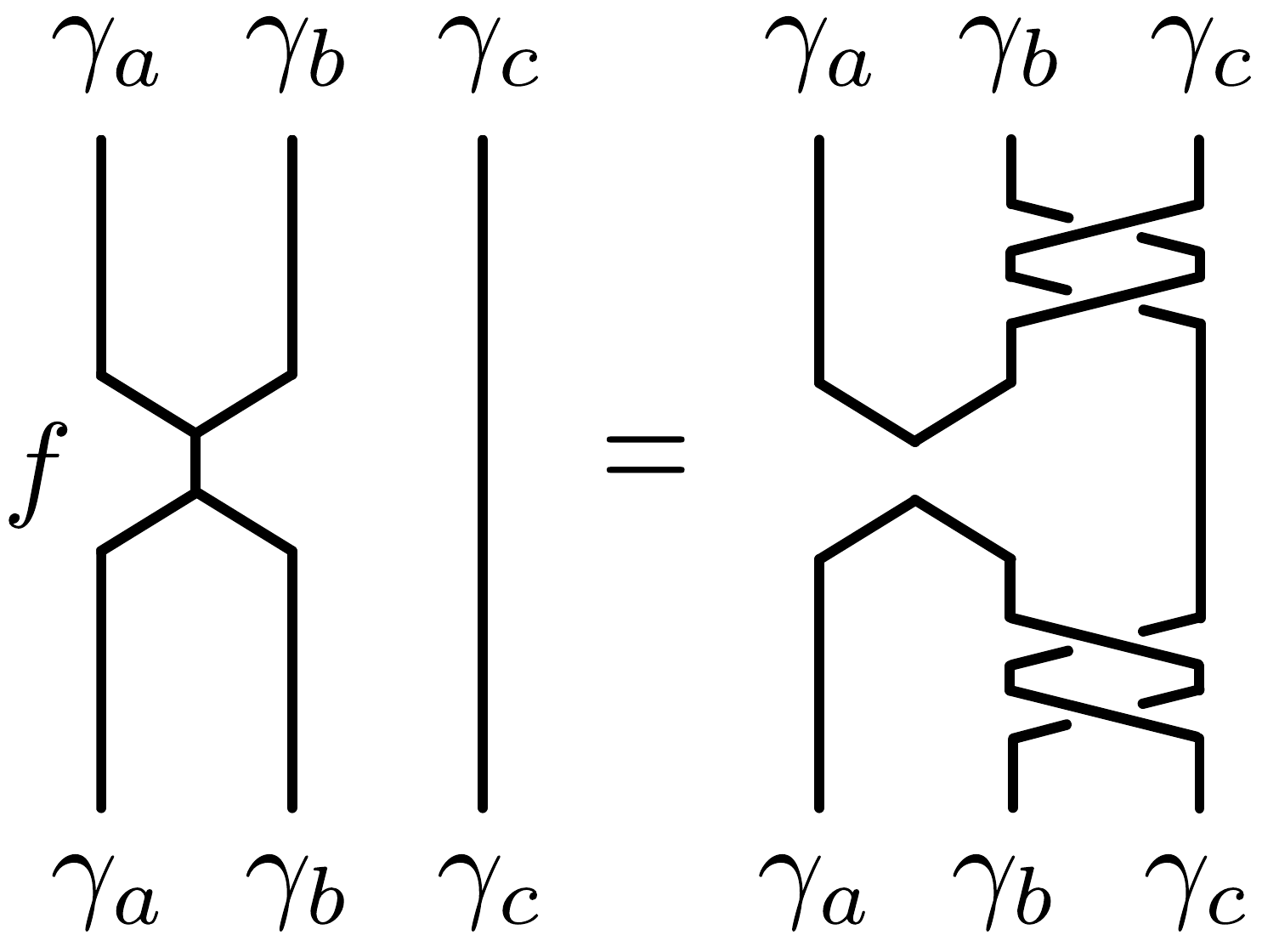},
        \end{equation} &
        \begin{equation}
            \label{Tran-4}
            \includegraphics[width=0.35\columnwidth]{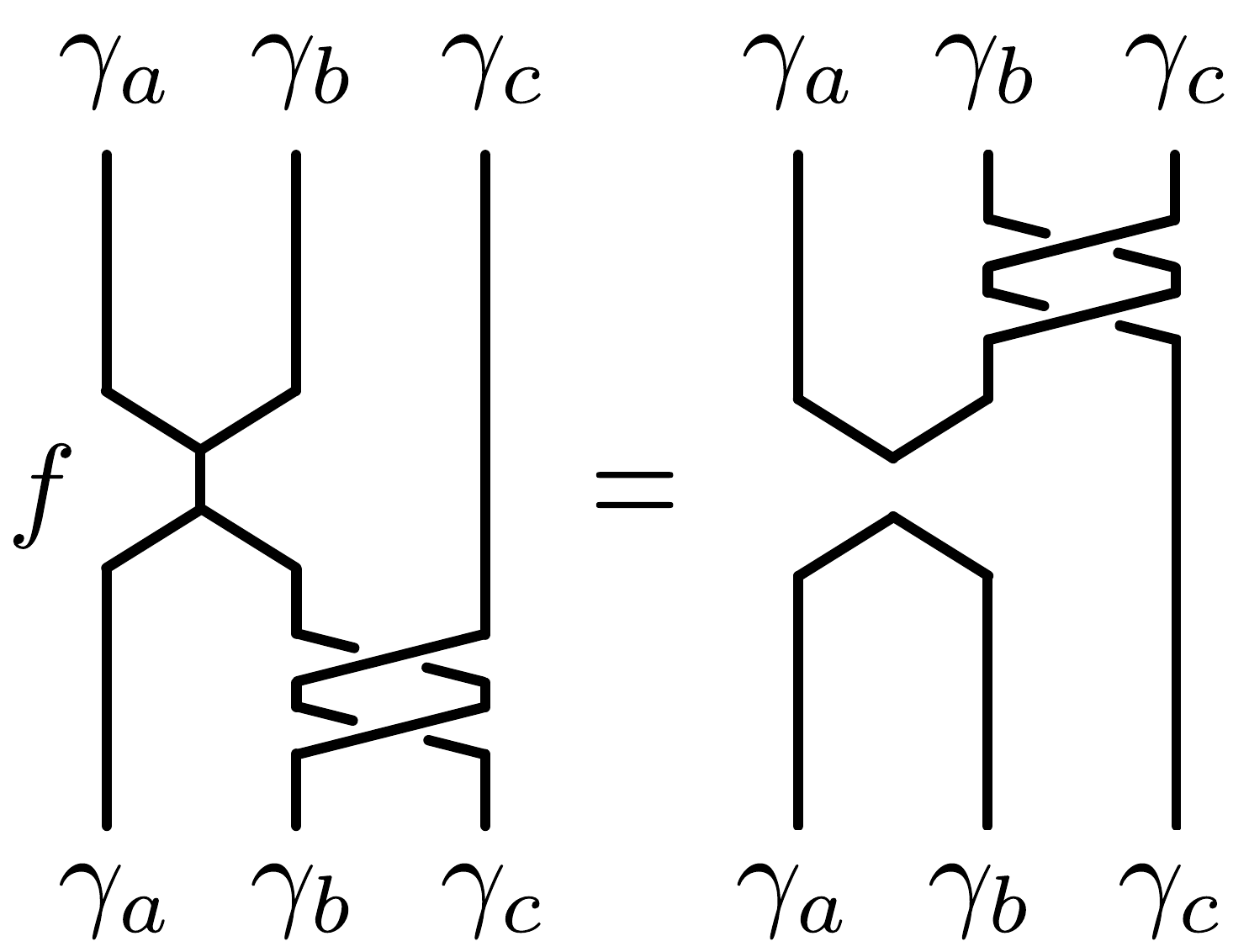}.
        \end{equation}
    \end{tabularx}
\end{subequations}
We find that the transformation of undesired outcomes to desired outcomes using Eq.~\eqref{Tran-3} and Eq.~\eqref{Tran-4} within the framework of full isotopy formalism can be achieved in different ways: the former adds the additional braiding operations as the payment; the latter transports the exchange operations occurring before the measurement to after the measurement. By combining these two equations and the full isotopy formalism, we can convert all the fermion measurement results to the vacuum while transferring the additional braiding operations after the final measurement. In Fig.~\ref{Correction}(a), we demonstrate how to propagate undesired outcomes, taking the example of an undesired outcome at the first measurement of the CZ gate (i.e.$\Pi_{5678}^{+} \Pi_{35}^{+} \Pi_{34}^{+} \Pi_{46}^{-} \Pi_{5678}^{+}$, detailed deformation version available in \cite{supp}).
 Taking Eq.~\eqref{Tran-3} and Eq.~\eqref{Tran-4} (dashed gray and blue circles) in succession, the undesired outcome shifts along the time axis from the first to the third measurement (the third diagram in Fig.~\ref{Correction}(a)). Repeating the similar procedure with Eq.~\eqref{Tran-3}, the consequences with undesired measurement outcome is transformed to the desired one with additional full braiding operations, which ends up with a full braiding of $\gamma_3$ and $\gamma_4$. Therefore, we only need to eliminate the full braiding by imposing an opposite braiding. Fortunately, the full braiding of MZMs can be achieved by evolving the MZMs in the parameter space without moving them. Therefore, we can correct the undesired outcome without either moving MZMs or through the measurements. 

To demonstrate the correction procedure more physically, we consider a specific hybrid system consisting of a superconductor (SC), a two-dimensional topological insulator (2DTI), and a ferromagnetic insulator (FI)  (Fig.~\ref{Correction}(b)). The low energy BdG Hamiltonian in spinor basis $\hat{c}(\bm{r})=\bigr[c_{\uparrow}(\bm{r}),c_{\downarrow}(\bm{r}),c^{\dagger}_{\downarrow}(\bm{r}),-c^{\dagger}_{\uparrow}(\bm{r})\bigr]^T$ takes
    \begin{eqnarray}
    \label{Ham-1}
    \hat{H}=\left(\begin{array}{cc}
    		h(\hat{p})+\bm{m}(\bm{r})\cdot\sigma&\Delta_{\rm SC}(\bm{r}) e^{-i\varphi(\vec{r})}\\
    		\Delta_{\rm SC}(\bm{r})e^{i\varphi(\vec{r})}  &-h(\hat{p})+\bm{m}(\bm{r})\cdot\sigma
    	\end{array}\right),
    \end{eqnarray}
where $\sigma_{x,y,z}$ are Pauli matrices in spin space, $h(\bm{\hat{p}})=v_{\rm f} \hat{p}\sigma_{z}-\mu$, with $v_{\rm f}$ the Fermi velocity of edge states and $\mu$ the chemical potential, $\Delta(\bm{r})$ is the proximity induced $s$-wave SC gap amplitude with SC phase $\varphi(\vec{r})$ and $\bm{m}(\bm{r})$ is the proximity induced exchange field. Here $\varphi(\vec{r})$ is the superconducting phase between each SC island and the outer SC. As the superconductivity and the ferromagnet open topologically equivalent gaps, each SC/FI interface supports one MZM. Taking $\gamma_1$ and $\gamma_{2}$ in Fig.~\ref{Correction}(b) for example, given the SC phase $\varphi_{1,2}$ and the magnetization direction $(\theta,\alpha)$ on a $S^2$ sphere (Fig.~\ref{Correction}(c)).

In the context of Majorana evolution involving magnetization winding, it is crucial to note that when the direction of magnetization $\bm m$ completes a closed trajectory enclosing the spin-orbit ($z$) axis, the Majorana wave functions in both SC and FI regions span $2\pi$ solid angle in the Bloch sphere (Fig.~\ref{Correction}(c)). This results in a $\pi$ monodromy phase, corresponding to a full braiding operation of $\exp(-\pi \gamma_1 \gamma_2 /2)$ \cite{Alicea2012}. A key observation is that the quantization of the monodromy phase is homotopy to the winding number of the magnetization around the SOC axis. Therefore, implementing braiding MZMs by varying the magnetization is topologically protected. Similarly, winding the superconducting phase also introduces the quantized monodromy phase $n\pi$ into the Majorana wave function \cite{supp}. Therefore changing the magnetic field direction by 0-$2\pi$ and changing the superconducting phase by 0-$2\pi$, the evolution of Majorana spin accumulates the same phase in spin and charge parameter spaces, respectively, which is equivalent to performing two successive exchanges of the corresponding MZMs, as shown in Fig.~\ref{Correction}(d).  

  {\it Conclusion-} We have proposed an ancilla-free measurement-based scheme to implement Clifford quantum gates with full topological protection. This design enables us to identify the minimal measurement sequences and allows the systematic construction of Clifford gates. Additionally, the deterministic quantum gates can be achieved through fully braiding Majorana modes in parameter space with topological protection. Our study provides valuable insight into the optimal design for topological quantum computation.

\begin{acknowledgments}
\section*{Acknowledge}
    We acknowledge useful discussions with Zheng-Xin Liu, Dong-Ling Deng, and Yue Yu. X. Liu acknowledges the support by the Innovation Program for Quantum Science and Technology (Grant No. 2021ZD0302700) and the National Natural Science Foundation of China (NSFC) (Grant No.12074133). X. J. Liu acknowledges the support by the Innovation Program for Quantum Science and Technology (Grant No. 2021ZD0302000), the National Natural Science Foundation of China (NSFC) (Grant No. 11825401 and No. 11921005), National Key Research and Development Program of China (2021YFA1400900), and the Strategic Priority Research Program of the Chinese Academy of Science (Grant No. XDB28000000). 
\end{acknowledgments}

\bibliography{Main.bbl}

\clearpage
\begin{appendix}
\begin{widetext}
\begin{center}
\begin{Large}
\textbf{Supplemental Materials}
\end{Large}
\end{center}

\section{Correcting the undesired outcomes in detail}
We demonstrate in detail two scenarios that are sufficient to handle the occurrence of other undesired outcomes in our measurement sequence. The first one is an exhaustive version of Eq.~\eqref{Cor1}, i.e.$\Pi_{5678}^{+} \Pi_{35}^{+}  \Pi_{34}^{+} \Pi_{46}^{-} \Pi_{5678}^{+}$:
    \beqn\label{Cor1}
    \includegraphics[width= 14 cm]{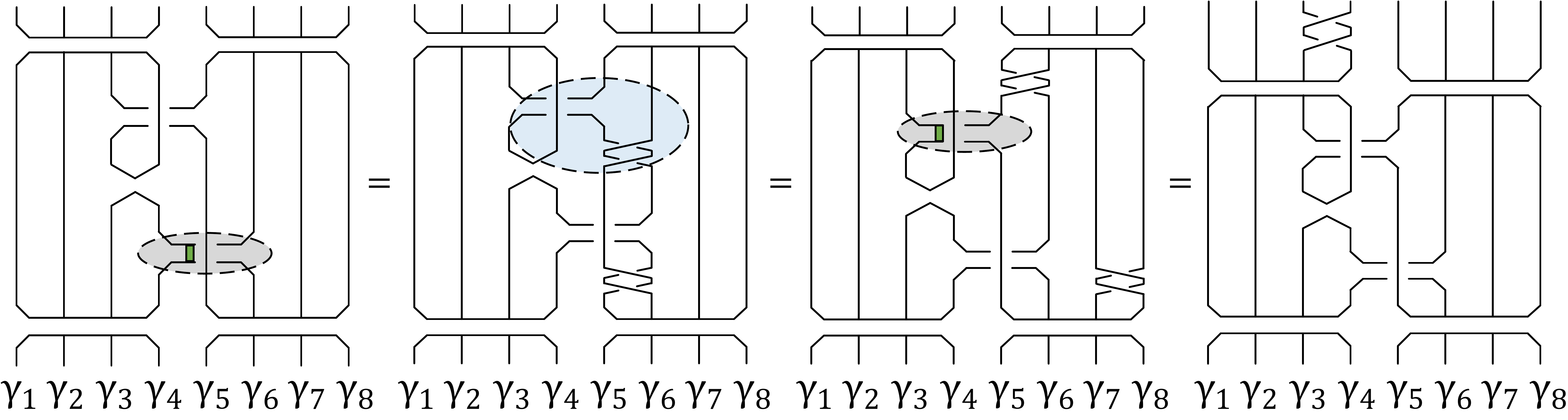}.
    \eeqn
Taking Eq.~\eqref{Tran-3} and Eq.~\eqref{Tran-4} (dashed gray and blue circles) in succession, the consequences with undesired measurement outcome is transformed to the desired one with additional full braiding operations, which ends up with a full braiding of $\gamma_3$ and $\gamma_4$.

The second scenario is the second measurement with non-expected results, i.e.$\Pi_{5678}^{+} \Pi_{35}^{+} \Pi_{34}^{-} \Pi_{46}^{+} \Pi_{5678}^{+}$:
    \beqn
    \includegraphics[width= 14 cm]{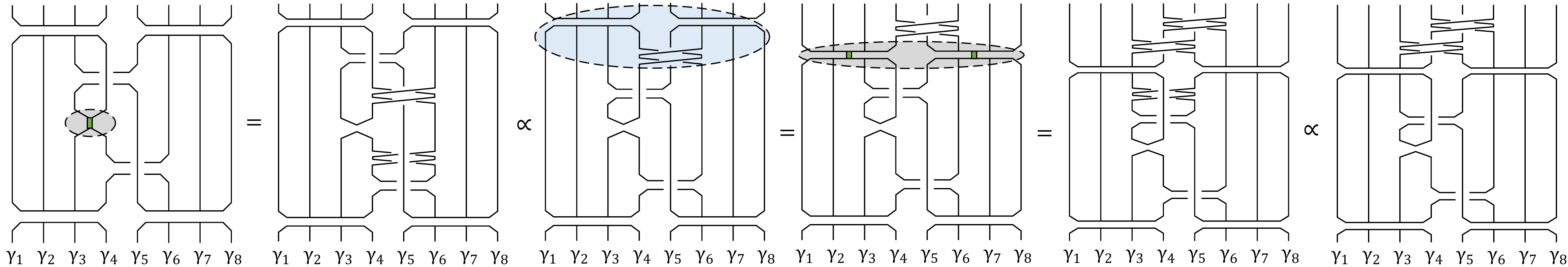}.
    \eeqn
The corresponding manipulation we gain is the combination of a full braiding of $\gamma_3$ and $\gamma_5$ and a full braiding of $\gamma_4$ and $\gamma_6$. Other scenarios can be handled in this way to identify the corresponding manipulations that allow for deterministic implementation of the CZ gate.

\section{Measurement sequences of CZ-gate}
\begin{table}[htbp]
 \centering
  \caption{Measurement sequences of CZ-gate \label{tab-1}}

\begin{tabular}{|cc|}
\hline
\multicolumn{1}{|c|}{$(\Pi_{1234}^{+}\Pi_{5678}^{+})\rightarrow \Pi_{15}\rightarrow \Pi_{12}\rightarrow \Pi_{26}\rightarrow \Pi_{5678}$} & $(\Pi_{1234}^{+}\Pi_{5678}^{+})\rightarrow \Pi_{26}\rightarrow \Pi_{12}\rightarrow \Pi_{15}\rightarrow \Pi_{5678}$ \\ \hline
\multicolumn{1}{|c|}{$(\Pi_{1234}^{+}\Pi_{5678}^{+})\rightarrow \Pi_{15}\rightarrow \Pi_{56}\rightarrow \Pi_{26}\rightarrow \Pi_{5678}$} & $(\Pi_{1234}^{+}\Pi_{5678}^{+})\rightarrow \Pi_{26}\rightarrow \Pi_{56}\rightarrow \Pi_{15}\rightarrow \Pi_{5678}$  \\ \hline
\multicolumn{1}{|c|}{$(\Pi_{1234}^{+}\Pi_{5678}^{+})\rightarrow \Pi_{17}\rightarrow \Pi_{12}\rightarrow \Pi_{28}\rightarrow \Pi_{5678}$} &                   $(\Pi_{1234}^{+}\Pi_{5678}^{+})\rightarrow \Pi_{28}\rightarrow \Pi_{12}\rightarrow \Pi_{17}\rightarrow \Pi_{5678}$ \\ \hline
\multicolumn{1}{|c|}{$(\Pi_{1234}^{+}\Pi_{5678}^{+})\rightarrow \Pi_{17}\rightarrow \Pi_{78}\rightarrow \Pi_{28}\rightarrow \Pi_{5678}$} &                   $(\Pi_{1234}^{+}\Pi_{5678}^{+})\rightarrow \Pi_{28}\rightarrow \Pi_{78}\rightarrow \Pi_{17}\rightarrow \Pi_{5678}$ \\ \hline
\multicolumn{1}{|c|}{$(\Pi_{1234}^{+}\Pi_{5678}^{+})\rightarrow \Pi_{35}\rightarrow \Pi_{34}\rightarrow \Pi_{46}\rightarrow \Pi_{5678}$} &                   $(\Pi_{1234}^{+}\Pi_{5678}^{+})\rightarrow \Pi_{46}\rightarrow \Pi_{34}\rightarrow \Pi_{35}\rightarrow \Pi_{5678}$  \\ \hline
\multicolumn{1}{|c|}{$(\Pi_{1234}^{+}\Pi_{5678}^{+})\rightarrow \Pi_{35}\rightarrow \Pi_{56}\rightarrow \Pi_{46}\rightarrow \Pi_{5678}$} &                   $(\Pi_{1234}^{+}\Pi_{5678}^{+})\rightarrow \Pi_{46}\rightarrow \Pi_{56}\rightarrow \Pi_{35}\rightarrow \Pi_{5678}$  \\ \hline
\multicolumn{1}{|c|}{$(\Pi_{1234}^{+}\Pi_{5678}^{+})\rightarrow \Pi_{37}\rightarrow \Pi_{34}\rightarrow \Pi_{48}\rightarrow \Pi_{5678}$} &                   $(\Pi_{1234}^{+}\Pi_{5678}^{+})\rightarrow \Pi_{48}\rightarrow \Pi_{34}\rightarrow \Pi_{37}\rightarrow \Pi_{5678}$   \\ \hline
\multicolumn{1}{|c|}{$(\Pi_{1234}^{+}\Pi_{5678}^{+})\rightarrow \Pi_{37}\rightarrow \Pi_{78}\rightarrow \Pi_{48}\rightarrow \Pi_{5678}$} &                   $(\Pi_{1234}^{+}\Pi_{5678}^{+})\rightarrow \Pi_{48}\rightarrow \Pi_{78}\rightarrow \Pi_{37}\rightarrow \Pi_{5678}$   \\ \hline
\end{tabular}
\end{table}

\section{Associated with other two qubit Clifford gates}
To demonstrate the universality of the spirit of geometric visualization, we additionally show measurement sequences for two 2-qubit Clifford gates of broad interest: CX(Y)-gate and iCZ-gate.
\subsection{1.CX(Y)-gate}
The previous exploration of the measurement scheme for CZ gates was completed using the Pauli matrix rewritten into pairs of MZMs for $i\gamma_{1}\gamma_{2} = s_{z}$, $ i\gamma_{7}\gamma_{8} = \tau_{z}$ and $ -\gamma_{5}\gamma_{6}\gamma_{7}\gamma_{8} = \sigma_{z}$. The measurement sequence of the obtained CZ gate is $\Pi_{1234}^{(+1)}\Pi_{5678}^{(+1)}(s_0\tau_0\sigma_z)\rightarrow \Pi_{46}^{(+1)}(s_0\tau_0\sigma_x)\rightarrow \Pi_{34}^{(+1)}(s_z\tau_0\sigma_z)\rightarrow \Pi_{35}^{(+1)}(-s_z\tau_z\sigma_x)\rightarrow \Pi_{5678}^{(+1)}(s_0\tau_0\sigma_z)$. Hence we gain $i\gamma_6\gamma_7 = s_0\tau_x\sigma_z $ and $i\gamma_5\gamma_8 = s_0\tau_x\sigma_z$.

It is known that CZ-gate and CX-gate can be related by the unitary transformation, $CX = (I \bigotimes U^{\dagger})CZ(I\bigotimes U)$, in which U is a unitary transformation. Gate base is a 4d-subspace belonging to the workspace $\sigma_z = +1$ and the auxiliary space is the part of $\sigma_z = -1$, denoted by $s_z\tau_z$ and $- s_z\tau_z$, respectively. Meanwhile the unitary transformation transform $i\gamma_5\gamma_6 = s_0\tau_z\sigma_z $ into $i\gamma_6\gamma_7 = s_0\tau_x\sigma_z $, which can be expressed as $U_{567} = \frac{1+\gamma_5\gamma_6}{\sqrt{2}}\frac{1+\gamma_6\gamma_7}{\sqrt{2}} = \frac{I-is_0\tau_z\sigma_z-is_0\tau_x\sigma_z-is_0\tau_y\sigma_0}{2}$. For gate base, $U_{567}$ works as $\frac{I-is_0\tau_z-is_0\tau_x-is_0\tau_y}{2}$ which converts CZ-gate to CX-gate. Then we have the corresponding CX-gate measurement sequence as $\Pi_{1234}^{(+1)}\Pi_{5678}^{(+1)}\rightarrow \Pi_{47}^{(+1)}\rightarrow \Pi_{34}^{(+1)}\rightarrow \Pi_{36}^{(+1)}\rightarrow \Pi_{5678}^{(+1)}$. 

Similarly, we consider the transformation of $i\gamma_5\gamma_6 = s_0\tau_z\sigma_z $ into $i\gamma_5\gamma_7 = s_0\tau_y\sigma_z $, which can be expressed as $U_{67} = \frac{1+\gamma_6\gamma_7}{\sqrt{2}} = \frac{1-is_0\tau_x\sigma_z}{\sqrt{2}}$. Thus we have the corresponding sequence of CY-gates $\Pi_{1234}^{(+1)}\Pi_{5678}^{(+1)}\rightarrow \Pi_{47}^{(+1)}\rightarrow \Pi_{34}^{(+1)}\rightarrow \Pi_{35}^{(+1)}\rightarrow \Pi_{5678}^{(+1)}$.

\subsection{2.iCZ-gate}
The approach mentioned in the main text helps us to decipher the iCZ sequence visually, where iCZ gate is
$\begin{pmatrix}
1 & 0 & 0 & 0 \\
0 & 1 & 0 & 0 \\
0 & 0 & i & 0 \\
0 & 0 & 0 & -i \\
\end{pmatrix}$
It is considered that the state $|1111\rangle$ needs to obtain a $-\pi/2$ phase, which can be provided by the closed path 
\beqn
\ket{+z} \rightarrow \ket{+x} \rightarrow \ket{-z} \rightarrow \ket{-y} \rightarrow \ket{+z}.
\eeqn
of the spherical axis. Without difficulty, there are only the following eight sequences that implement iCZ gate.

\begin{table}[htbp]
 \centering
  \caption{Measurement sequences of iCZ-gate \label{tab-2}}

\begin{tabular}{|cc|}
\hline
\multicolumn{1}{|c|}{$(\Pi_{1234}^{+}\Pi_{5678}^{+})\rightarrow \Pi_{15}\rightarrow \Pi_{12}\rightarrow \Pi_{16}\rightarrow \Pi_{5678}$} & $(\Pi_{1234}^{+}\Pi_{5678}^{+})\rightarrow \Pi_{17}\rightarrow \Pi_{12}\rightarrow \Pi_{18}\rightarrow \Pi_{5678}$ \\ \hline
\multicolumn{1}{|c|}{$(\Pi_{1234}^{+}\Pi_{5678}^{+})\rightarrow \Pi_{25}\rightarrow \Pi_{12}\rightarrow \Pi_{26}\rightarrow \Pi_{5678}$} & $(\Pi_{1234}^{+}\Pi_{5678}^{+})\rightarrow \Pi_{27}\rightarrow \Pi_{12}\rightarrow \Pi_{28}\rightarrow \Pi_{5678}$  \\ \hline
\multicolumn{1}{|c|}{$(\Pi_{1234}^{+}\Pi_{5678}^{+})\rightarrow \Pi_{35}\rightarrow \Pi_{34}\rightarrow \Pi_{36}\rightarrow \Pi_{5678}$} &                   $(\Pi_{1234}^{+}\Pi_{5678}^{+})\rightarrow \Pi_{37}\rightarrow \Pi_{34}\rightarrow \Pi_{38}\rightarrow \Pi_{5678}$ \\ \hline
\multicolumn{1}{|c|}{$(\Pi_{1234}^{+}\Pi_{5678}^{+})\rightarrow \Pi_{45}\rightarrow \Pi_{34}\rightarrow \Pi_{46}\rightarrow \Pi_{5678}$} &                   $(\Pi_{1234}^{+}\Pi_{5678}^{+})\rightarrow \Pi_{47}\rightarrow \Pi_{34}\rightarrow \Pi_{48}\rightarrow \Pi_{5678}$ \\ \hline

\end{tabular}
\end{table}

The existence of another closed path of the state $|1111\rangle$ on the observation possibly forming iCZ:
\beqn
\ket{+z} \rightarrow \ket{+x} \rightarrow \ket{-y} \rightarrow \ket{-x} \rightarrow \ket{+z}.
\eeqn
, however, no counterpart to the 2-majorana sequence of measurement operators could be found.

\section{Majorana wave function}
Considering  $\text{FI-SC}$ junction proximity on the edge states of 2DTI,  $h(\hat{p})$ and $\bm{m}(\bm{r})\cdot\sigma$ in Eq.\ref{Ham-1} can be reduced to $v_{\rm f}\hat{p}\sigma_z$ and $(m_{\parallel}(\bm r)e^{i\alpha}+m_z(\bm r))\cdot\sigma$ respectively, where $m_z=|\bm m|\sin\theta$. The position vector $\bm{r}$  is expressed as $\bm{r}=x\bm{e_{\rm x}}$ in this one dimension model. The definitions of the magnetic configuration,  chemical potential and superconducting  pairing potential in real space take
\beqn
\label{sc}
\Delta_{\text{SC}}(x)=\Delta\Theta(x), m_{\parallel/{\rm z}}(x)=m_{\parallel/{\rm z}}\Theta(-x),
\mu(x)=\mu_{\text{SC}}\Theta(x)+\mu_{\text{FI}}\Theta(-x).
\eeqn
The wave function for the electron and hole band in FI region($x<0$) and SC region($x>0$) are  straightforward to show that
\beqn
&&\Psi_{\text{FI}}^{\rm e}(x)=(v_{\rm f}p+m_{{\rm z}}+\mu_{\text{FI}}+E,m_{\parallel}e^{i\alpha},0,0)e^{i{k}_{\text{FI}}^{\rm e}x}, \Psi_{\text{FI}}^{\rm h}(x)=(0,0,-v_{\rm f}p+m_{\rm z}-\mu_{\text{FI}}+E,m_{\parallel}e^{i\alpha})e^{i{k}_{\text{FI}}^{\rm h}x}(x<0),\nonumber\\
&&\Psi_{\text{SC}}^{\rm e}(x)=(\frac{v_{\rm f}p-\mu_{\text{SC}}+E}{\Delta},0,1,0)e^{i{k}_{\text{SC}}^{\rm e}x}, \Psi_{\text{SC}}^{\rm h}(x)=(0,\frac{-v_{\rm f} p-\mu_{\text{SC}}+E}{\Delta},0,1)e^{i{k}_{\text{SC}}^{h}x}(x>0),
\eeqn
 where wave vectors $k_{\text{FI/SC}}^{e/h}$ are defined as
\beqn
 {k}_{\text{FI}}^{\rm {e/h}}=\frac{- i\sqrt{m_{\parallel}^2-(E\pm\mu_{\text{FI}})^2}\mp m_{{\rm z}}}{\hbar v_{\rm f}},{k}_{\text{SC}}^{\rm {e/h}}=\frac{i\sqrt{\Delta^2-E^2}\pm\mu_{\text{SC}}}{\hbar v_{\rm f}}.
\eeqn

Considering the zero energy solution of this BDG Hamiltonian, the wave functions in FI and SC region respectively can be written as
\beqn
&&\Psi_{\text{FI}}(x)=a_{\rm e}(e^{-ik_{\rm m}x},e^{i(\alpha+\varphi)}e^{-ik_{\rm m}x},0,0)+a_{\rm e}^{\ast}(0,0,e^{-i(\alpha+\varphi)}e^{ik_{\rm m}x},-e^{ik_{\rm m}x})^{T}e^{k_{\text{FI}}x}(x<0),\nonumber\\
&&\Psi_{\text{SC}}(x)=m_{\rm e}(ie^{ik_{\text{sc}}x},0,e^{ik_{\text{sc}}x},0)+m_{\rm e}^{\ast}(0,e^{-ik_{\text{sc}}x},0,ie^{-ik_{\text{sc}}x})^{T}e^{-K_{\text{SC}}x}(x>0),
\eeqn
where these parameters $\varphi, k_{\text{FI}}, k_{\rm m}, K_{\text{SC}},k_{\text{sc}} $ are defined for simplification as
\beqn
e^{i\varphi}=\frac{i\sqrt{m_{\parallel}^2-\mu_{\text{FI}}^2}+\mu_{\text{FI}}}{m_{\parallel}}, k_{\text{FI}}=\frac{\sqrt{m_{\parallel}^2-\mu_{\text{FI}}^2}}{\hbar v_{\rm f}}, K_{\text{SC}}=\frac{\Delta}{\hbar v_{\rm f}}, k_{\text{m/sc}}=\frac{m_{\rm z}/\mu_{\text{SC}}}{\hbar v_{\rm f}}.
\eeqn
Coefficients $a_{\rm e}, m_{\rm e}$ are determined by matching the boundary condition $\Psi_{\text{FI}}(0)=\Psi_{\text{SC}}(0)$. The results of this zero energy wave functions take
\beqn
&\Psi_{\text{FI/SC}}(x)=(\psi_{\text{FI/SC}}^{\rm e}(x), i\sigma_y \psi_{\text{FI/SC}}^{\ast\rm e}(x))^{\rm T},\nonumber\\
&\psi_{\text{FI}}^{\rm e}=e^{i(\frac{\pi}{4}-k_{\rm m}x)}(e^{-i\frac{\phi}{2}},e^{i\frac{\phi}{2}}),\psi_{\text{SC}}^{\rm e}=e^{i\frac{\pi}{4}}(e^{-i\frac{\phi}{2}+ik_{\text{sc}}x},e^{i\frac{\phi}{2}-ik_{\text{sc}}x}),
\eeqn
where $\phi=\alpha+\varphi$.

\end{widetext}
\end{appendix}
\end{document}